# OPERATIONAL USE OF IONIZATION PROFILE MONITORS IN THE FERMILAB MAIN INJECTOR*

Denton K. Morris, Philip Adamson, David Capista, Ioanis Kourbanis, Thomas Meyer, Kiyomi Seiya, David Slimmer, Ming-Jen Yang, James Zagel, Fermilab, Batavia, IL 60510-5011, U.S.A.

*Abstract*

Ionization profile monitors (IPMs) are used in the Fermilab Main Injector (MI) to monitor injection lattice matching by measuring turn-by-turn sigmas at injection and to measure transverse emittance of the beam during the acceleration cycle. The IPMs provide a periodic, non-destructive means of performing turn-by-turn emittance measurements where other techniques are not applicable. As Fermilab is refocusing its attention on the intensity frontier, non-intercepting diagnostics such as IPMs are expected to become even more important. This paper gives an overview of the operational use of IPMs for emittance measurements and injection lattice matching measurements at Fermilab, and summarizes the future plans.

## INSTALLATION CONFIGURATIONS

The Main Injector currently has three IPMs. Two of the IPMs are electrostatic units, one horizontal and one vertical, operating at 28KV, collecting ions and producing data every 11.1 µSec. The third IPM is a horizontal unit which is installed within a 1 KGauss magnetic field and a 10 KV clearing field which allows the collection of electrons[1]. The magnetic field confines liberated electrons to orbits smaller than the anode pick-up strips which minimizes the deleterious effect of space charge from the beam. The Main Injector IPMs are limited to 65K samples by the current digitizer; however, by skipping turns during acquisition, the full cycle of beam can be measured.

## OPERATING CONDITIONS

The Fermilab Main Injector is a very dynamic machine operating over a large range of intensities. Antiproton transfers from the Antiproton Accumulator to the Recycler via the Main Injector have intensities of $5 \times 10^{10}$ to $2 \times 10^{11}$ antiprotons in four 2.5 MHz bunches ($1.2 \times 10^{10}$ to $5 \times 10^{10}$ particles per bunch) at an energy of 8 GeV. Protons accelerated from 8 GeV to 120 GeV for Antiproton production and NUMI targeting will have total intensities of $4.6 \times 10^{13}$ protons in 6 groups of 84 bunches and peak single bunch intensities of $1 \times 10^{11}$ protons.

Combined antiproton production and NUMI targeting cycles have a repetition rate of 2.2 seconds with beam in the machine for 1.5 seconds of the cycle. Peak power in the Main Injector for these cycles is 400 kWatts.

Vacuum in the IPM region is maintained at $1 \times 10^{-8}$ Torr.

## CONTROLS INTERFACE

LabVIEW was chosen as the environment for the front end program because of its facilities for tying together different types of hardware (GPIB, VME, Ethernet) and software (Accelerator Controls NETwork (ACNET), dlls). During normal operations the systems are controlled through ACNET, but the front-end program provides complete functionality for running the systems locally. Beam measurements are configured with a set of file based specifications that include trigger and event types, timing delays, high voltage settings, analysis parameters and data logging preference. Measurements are initiated by activating a specification number through either the LabVIEW or ACNET control systems [2].

## ACNET INTERFACE

The IPM systems are most often used through the ACNET controls system. There is a dedicated application for generating measurement specifications and for displaying data collected on the front end. There is also a front end state to synchronize data collection timing. Measurement specifications can be activated manually but are most often enacted by the accelerator Sequencer application which coordinates Antiproton transfers and colliding beams setups. Based on state changes indicating new data has been acquired and processed, a central data logger will archive the data presented by the front end to ACNET.

## MEASUREMENT CAPABITITY

The IPMs in the Main Injector integrates a 1.6 µSec batch of 84 bunches to generate a beam profile every 11.1 µSec, corresponding to each revolution of the beam. Each profile consists of 60 channels from the micro-channel plate, spaced 0.5mm apart. The IPMs are able to capture 65,000 turns of data covering a continuous 0.72 seconds of beam. Cycles longer than 0.72 seconds can be measured by skipping one or more turns as needed. Specific periods within a cycle can be studied by delaying the start of the acquisition.

The common automated use is to start data collection at injection and collect 500 turns of data to be logged and analyzed. The data for each turn is fit to a Gaussian profile which is used to generate parameterized values for the both the beam position and the sigma. In addition, the amplitude of initial oscillations, frequency of the oscillation, phase of the oscillation and the decay rate of any oscillations are recorded for both positions and sigmas. Along with this primary data, the quality of the fits, uncertainty of the fits, signal-to-noise ratios, peak values, and general operating parameters are recorded for each data set. This allows us to monitor current beam dynamics, trends in beam performance and the quality of the IPM operations over time.

Figure 1 shows the basic data used to generated the fit parameters. In the data you can see a sigma oscillation and an injection position error.

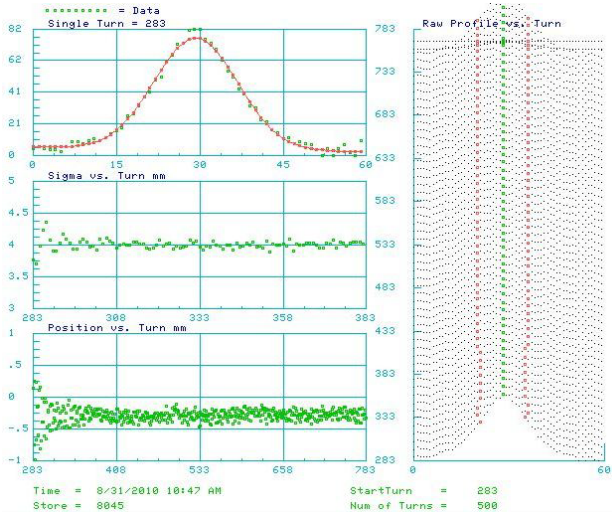

Figure 1: Measured injection errors from the MI-8 transfer line. Top left is a fit to a single turn 60 channel data sample. Channel separation is 0.5mm. Middle left is the measured sigmas from 100 turns showing an initial oscillation damping down. Bottom left is the position error at injection. Right side shows the 500 profile samples averaged for ease of display.

The data collected from the IPM micro-channel plate is fit to the equation:

$$a_0 + a_1 e^{(-a_2 x)} \sin(2\pi x a_3 + a_4)$$

for both the sigma and position data. The values of a0 through a4 give the results:
 $a_0$ = signal offset
 $a_1$ = amplitude of oscillations
 $a_2$ = decay rate of oscillations
 $a_3$ = frequency of oscillations
 $a_4$ = phase of oscillations

## OPERATIONAL USES

The IPMs are configured through state machines to automatically measure antiproton transfers from the Antiproton Accumulator to the Main Injector and from the Recycler to the Main Injector. All data is logged in the ACNET Datalogger and collected into the Shot Data Analysis database for easy retrieval. The IPMs are also used to periodically check proton injections from the Booster into the Main Injector to assure lattice matching is acceptable.

The automated measurements and periodic measurements have successfully identified problems in systems that we were able to quickly rectify. An example of this was the identification of a quadrupole magnet error in the 8 GeV injection line. Figure 2 shows how changing the 8GeV line quadrupole current, based on IPM data, resulted in a reduced spot size on the NUMI target at 120 GeV.

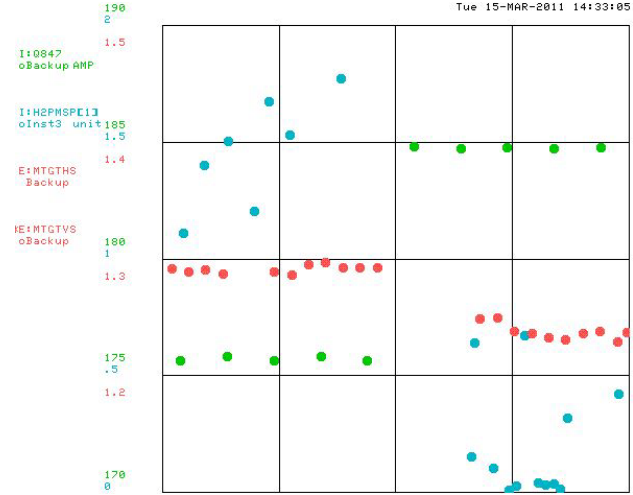

Figure 2: Increasing the current in a quadrupole (I:Q847, green points) by 9 amps reduced the oscillations of the beam sigma (cyan, value a1 in the equation) by 1mm. The result is a smaller spot size on the NUMI target (red points, product of the horizontal sigma and the vertical sigma at the NUMI target).

Similar measurements have resulted in changes that have increased the efficiency of antiproton transfers from the Antiproton Accumulator into the Recycler Ring via the Main Injector.

Figure 3 shows antiproton data from a week of transfers from the Accumulator and the Recycler into the Main Injector. The data shows the horizontal beam sigma (green) holding steady near 2mm for most of the transfers. Small increases in sigma occur when the Antiproton Source delivers larger than normal intensities of antiprotons that have undergone less transverse cooling. The isolated points in the 1mm range are the antiprotons that have been cooled in the Recycler Ring prior to transfer into the Main Injector for Tevatron colliding beam operations. Also shown are the horizontal position at injection (red), the frequency of the horizontal betatron oscillations (dark blue), and the amplitude of the horizontal sigma oscillations (cyan) near zero.

## OPERATIONAL LIFETIME AND MAINTANENCE

The gain of the micro-channel plate in the IPM degrades over time. The rate of degradation is dependent on both the voltage levels on the micro-channel plate and the integrated intensity of the beam the plate is exposed to when the voltage is on. The signal from the micro-channel plate can be optimized by adjusting the micro-channel plate voltage, but the higher the voltage the faster the plates are depleted.

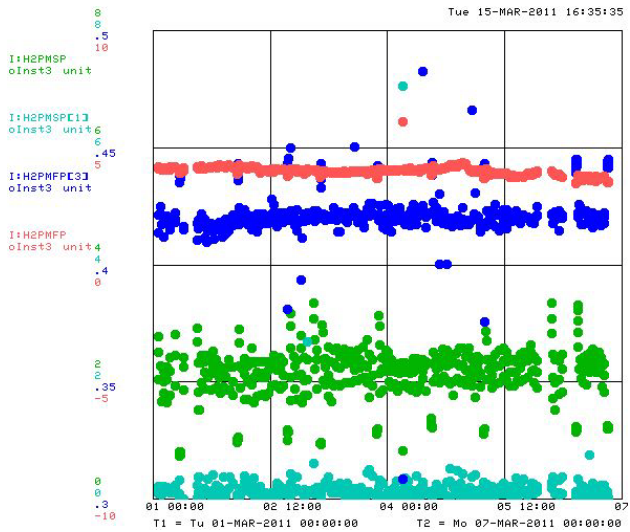

Figure 3: A week of IPM data showing beam sigma (green), oscillations of sigma (cyan), horizontal betatron oscillations (dark blue) and beam position (red).

To maximize the signal and the lifetime of the micro-channel plates the high voltage is monitored and set to achieve a peak of between 30 and 50 counts. This maximizes the signal to noise ratio and minimizes the uncertainty of the fit to the data. Lower voltage settings reduce the quality of the data and higher voltage settings unnecessarily degrade the plates at a quicker rate.

Running the micro-channel plate for an extended period produced the plate response degradation curve seen in Figure 4. Shown is the peak signal for measurements of constant intensity antiprotons using a fixed voltage on the micro-channel plate. The peak response is halved after exposure to 550 measurements of antiproton transfers from the Recycler Rings into the Main Injector. These measurements were taken over a period of 9 weeks. During this period of running the IPM was also exposed periodically to other beam cycles at a rate of approximately $5.4 \times 10^{14}$ protons per hour with the micro-channel plate at high voltage.

As the response of the micro-channel plate decreases the voltage can be raised to increase the gain and maintain good signal quality. Another option is to remotely reposition the plates to an area that has suffered less depletion. This allows the IPM to be used functionally at these exposure rates for about six months before it is necessary to access the system and replace the plates.

The process of replacing the micro-channel plates requires accessing the Main Injector vacuum and letting up four half cells, approximately 20 meters. It takes roughly one hour to access and prepare the IPM for maintenance, 30 minutes to replace the micro-channel plates, and another 7 to 8 hours to recover vacuum to a level where we can resume low intensity beam, with high intensity beam following in a few hours. Total time from beam off to full intensity beam for a micro-channel plate replacement is approximately 10 to 12 hours. This is a process that can be scheduled every six months. Future plans include the installation of additional beam valves to reduce the length of beam pipe let up for accessing the IPM to approximately 2 meters which will greatly reduce the vacuum recovery time.

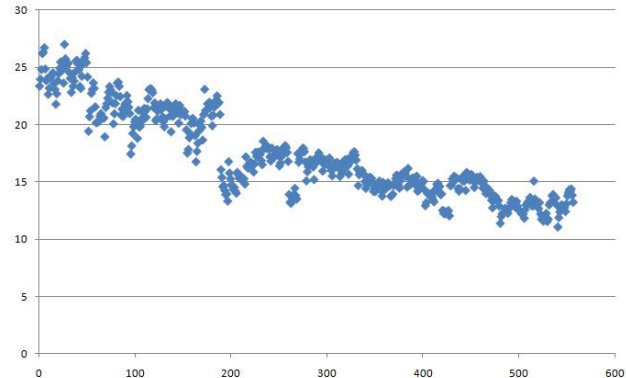

Figure 4: Micro-channel plate response peak degrading with exposure to beam. Peak channel counts versus data sample for a constant intensity and plate voltage.

## THE INTENSITY FRONTIER

The future of beam operations in the Main Injector is to push the intensity frontier for Neutrino physics. The expected number of protons injected per 120 GeV cycle is expected to increase slightly, from $4.6 \times 10^{13}$ up to $5.2 \times 10^{13}$ but the cycle length will be shortened from 2.2 seconds to 1.33 seconds. This will roughly double the beam power transmitted through the Main Injector at 120 GeV from 400 kWatts to 700 kWatts.

At the increased power levels the IPM system can expect to perform an automated measurement once every 6 minutes and experience the same signal degradation lifetime as current operations.